\documentclass[nofootinbib,preprintnumbers]{revtex4}
\usepackage{epsfig}
\usepackage{graphicx}
\usepackage{dcolumn}
\usepackage{bm}
\usepackage{amsmath}
\usepackage{amssymb}
\usepackage{multirow}

\begin{document}
\title{Fully hadronic decays of a singly produced vector-like top partner at the LHC}

\author{St\'ephanie Beauceron, Giacomo Cacciapaglia, Aldo Deandrea\footnote{also Institut Universitaire de France, 103 boulevard Saint-Michel, 75005 Paris, France}, Jos\'e D. Ruiz-\'Alvarez}
\affiliation{Universit\'e de Lyon, France; Universit\'e Lyon 1,
CNRS/IN2P3, UMR5822 IPNL, F-69622 Villeurbanne Cedex, France.}
\email[Email: ]{s.beauceron@ipnl.in2p3.fr, g.cacciapaglia@ipnl.in2p3.fr, deandrea@ipnl.in2p3.fr,j.ruiz-alvarez@ipnl.in2p3.fr}
\begin{abstract}
{Single production of vector-like top partners is becoming a major focus of new searches, as the increasing mass limits on vector-like 
quarks obtained by the ATLAS and CMS collaborations are such that single production is becoming competitive with respect to pair 
production. 
Typically searches focus on decays containing leptons in the final state in order to have less background from the Standard 
Model processes. However the current centre of mass energies available in the latest LHC runs limit the searches to low masses. Fully hadronic final 
states may be an alternative option for discovery, as they allow a larger number of signal events if backgrounds can be kept under 
control.
We study the fully hadronic decay of a singly produced vector-like top partner and we show a strategy to extract the signal over 
the background, considering as a benchmark the 20~fb$^{-1}$ of data collected at 8~TeV with the run I of the LHC.}
\end{abstract}

\preprint{LYCEN-2014-01}

\maketitle

\section{Introduction}

The discovery of the Higgs boson in 2012 allowed to establish the effective description of the fundamental laws of nature provided 
by the Standard Model of particle physics. The Large Hadron Collider (LHC) has also a strong potential for the discovery or exclusion of new particles, as 
expected in the different extensions of the Standard Model. Among these, many contain top partners (such as extra-dimensional 
models, little Higgs models, strong electroweak sector and composite Higgs models). These new multiplets are typically of vector-like
type, therefore vector-like quarks \cite{delAguila:1982fs} (VLQ) constitute a central piece of evidence for a set of models which 
summarize different possibilities of physics beyond the standard model. 
Many ATLAS \cite{Aad:2012uu,ATLAS:2013ima,Aad:2011yn,Aad:2012bt} and 
CMS \cite{CMSincl,Chatrchyan:2012af,Chatrchyan:2012fp,Chatrchyan:2012vu,CMS:2012ab,Chatrchyan:2011ay} searches have been 
recently performed in order to discover or set bounds on this type of particles related to the top sector. In many cases, as a simplifying assumption 
supported by theoretical expectations, only mixing to the third quark generation was considered. However partial mixing to the light 
generations should be allowed and is indeed a quite general expectation in the case of an extended CKM matrix (see  
\cite{Buchkremer:2013bha} for a general parameterisation and suggestions of LHC signatures of this type; for a discussion mainly 
focussing instead on the scenario of dominant third generation coupling to the vector-like quarks see \cite{Aguilar-Saavedra:2013qpa};
for a general discussion of mixing effects in the limit where the vector-like quarks are integrated out of the effective Lagrangian 
see \cite{delAguila:2000rc}). Experimentally these studies are now being performed in view of their relevance for phenomenology and model building constraints \cite{ATLAS:2012apa}, \cite{Garberson:2013jz}.
The crucial point is that even a 
small mixing to the first generation may have important phenomenological consequences as single production would be driven by the first 
generation couplings due to the large light partonic content in the colliding protons at the LHC. In such a scenario, the couplings to the third 
generation and in particular the top quark, would be very relevant in the decay modes, as experimentally these decays give 
several clear signatures with respect to background processes from the standard model. 
In this work, we consider the electroweak singly produced top partner (mainly through the flavour changing vertex $TZu$)
and the decay of the $T$ into a top quark and a Higgs boson~\footnote{The same final state has been considered in~\cite{Li:2013xba} in the case of exclusive couplings to the third generation.}. Finally, we considered hadronic decays of both the top (three jets) and the Higgs boson (two jets): we therefore designed our analysis on events with 5 final jets as signature, plus an additional forward jet from the production. 
This choice of fully hadronic final states has some important advantages with respect to other choices for the decays of the Higgs and top:
 in the first instance, larger branching ratios lead to a higher number of events. As a second point, we shall show that 
a full mass reconstruction is achievable, something not possible in the corresponding leptonic signatures.

\section{The model}
\label{sec:model}
From the theoretical side, our analysis is based on a general parameterisation \cite{Buchkremer:2013bha} of models with vector--like quarks. All different configurations basically depend on the choice of a type of multiplet for the new quarks (singlet, doublet, triplet, etc.) 
and the assignment of hypercharge. This work is focused on a non-standard doublet containing a top partner $T$ of electric 
charge  $2/3$ and an exotic X quark of electric charge $5/3$. This doublet corresponds to a shift of one unit of hypercharge with 
respect to a standard doublet and it is much less constrained by flavor than other choices because it does not contain a bottom quark partner. For a detailed analysis of 
the phenomenological implications for such a multiplet see \cite{Cacciapaglia:2010vn,Cacciapaglia:2011fx,Okada:2012gy}, while a collider study can be found in \cite{Cacciapaglia:2012dd,Azatov:2013hya}. 
From a model building point of view, this multiplet 
is also interesting as it emerges in some strong sector extensions of the Standard Model~\cite{Contino:2008hi,Mrazek:2009yu,Dissertori:2010ug,DeSimone:2012fs,Gopalakrishna:2013hua}. For this multiplet a large coupling of the top partner to 
the first generation of quarks is phenomenologically allowed~\cite{Cacciapaglia:2011fx}. Note however that our particular choice is not restrictive apart from the 
detailed values of branching ratio in the final state we analyse. In fact, a generic heavy $T$ will decay to standard model quarks and $W$, $Z$, $H$ bosons. 
As typical minimal masses allowed by the present LHC searches are in the 500-700~GeV range 
\cite{Chatrchyan:2012af,ATLAS:2012qe,ATLAS:2013ima,ATLAS:2013-051} from pair production analyses, 
to a good approximation the decay 
rates are consistent with the asymptotic predictions of the equivalence theorem (in the ratio $2:1:1$ for decays into $W$, $Z$, $H$ 
of a heavy top quark, but with different ratios if the quantum numbers are not the standard model ones). 
In the case of this non-standard doublet, the following decay modes are possible: $T\rightarrow Z u^{i}, \; H u^{i}$ in the ratio $1:1$, 
but depending on the selection of the couplings a given channel $i$ can be enhanced ($u^{i}$ means quarks of up type $i$). 
On the other hand, for the exotic partner $X\rightarrow  W^{+} u^{i}$ only, due to its exotic electric charge.

The effective Lagrangian for a general formulation of VLQ in \cite{Buchkremer:2013bha} can be restricted for $T$ and $X$ type of 
vector-like quarks in the non-standard doublet:
\begin{eqnarray}
  \mathcal{L} & = & \kappa_{T}\left\{ \sqrt{\frac{\zeta_{i}\xi_{Z}^{T}}{\Gamma_{Z}^{0}}}\frac{g}{2c_{W}}\left[ \bar{T}_{R}Z_{\mu}\gamma^{\mu}u^{i}_{R}\right]\right\} 
               -  \kappa_{T}\left\{ \sqrt{\frac{\zeta_{i}\xi_{H}^{T}}{\Gamma_{H}^{0}}}\frac{M}{v}\left[ \bar{T}_{L}Hu^{i}_{R}\right] + \sqrt{\frac{\zeta_{3}\xi_{H}^{T}}{\Gamma_{H}^{0}}}\frac{m_{t}}{v}\left[ \bar{T}_{R}Ht_{L}\right]\right\} \nonumber\\            
              & + & \kappa_{X}\left\{ \sqrt{\frac{\zeta_{i}}{\Gamma_{W}^{0}}}\frac{g}{\sqrt{2}}\left[ \bar{X}_{R}W^{+}_{\mu}\gamma^{\mu}u^{i}_{R}\right]\right\} +h.c.
\end{eqnarray} 
where we have neglected terms proportional to the light quark masses. The terms proportional to the top quark mass can be also 
neglected as they typically induce corrections of the order of 10\%-20\% for $M = 600$~GeV (therefore of the order of uncertainty in the 
present searches) except for the case of the coupling to the Higgs boson, which is therefore explicitly included. 
Apart from this chiral flipped term to the Higgs boson, all other terms have the same chirality. The Lagrangian terms with switched 
chirality (L $\leftrightarrow$ R) are possible, but suppressed as explained above and detailed in \cite{Buchkremer:2013bha}.
The parameters $\kappa_T$, $\kappa_X$ are the coupling strengths and determine the strength of single production. The parameters
$\zeta_{i}$ and $\xi_i$ are directly linked to the branching ratios ($\zeta_{jet}$ includes the contributions of the two light generations, which experimentally go both to jets):
 \begin{eqnarray} 
BR (T \to Z j) = \frac{\zeta_{jet} \xi^T_Z}{1+\zeta_3 \xi_H \delta_H}\,, & & BR (T \to Z t) = \frac{(1-\zeta_{jet}) \xi^T_Z}{1+\zeta_3 
\xi_H \delta_H}\,,\\
BR(T \to H j) = \frac{\zeta_{jet} (1-\xi^T_Z)}{1+\zeta_3 \xi_H \delta_H}\,, & &  BR(T \to H t) = \frac{(1-\zeta_{jet})
(1-\xi^T_Z) (1+\delta_H)}{1+\zeta_3 \xi_H \delta_H}\,, \nonumber\\
 & & \nonumber \\
BR(X \to W^+ j) = \zeta_{jet}\,, & & BR(X \to W^+ t) = (1-\zeta_{jet})\,.
 \end{eqnarray} 
The branching fractions of the top partner $T$ only depend on the two parameters 
$\zeta_{jet}$ and $\xi_Z^{T}$, as $\delta_H$ is a calculable function of the heavy mass scale for the vector-like quarks 
$M$ \cite{Buchkremer:2013bha}:
\begin{eqnarray} 
\delta_H  \sim 5 \frac{m^2_t}{M^2}\,, \label{eq:deltaH}
\end{eqnarray} 
where we kept the leading order in $1/M^2$. Note that in these formulae we kept the decay rates into $Z$ and $H$ arbitrary (parameterised by $\xi_Z^{T}$, which is equal to 1/2 for the non-standard doublet case). The X partner in the non-standard doublet has a branching which depends on a single 
parameter $\zeta_{jet}$. These approximate but quite robust results allow to describe easily the phenomenology of the non-standard 
doublet and its single production. For a top partner coming from other multiplets the situation is not much different in terms of decay 
modes, as only the extra decay to $W$ and quarks will be present, but those to $Z$, $H$ and quarks which we discuss in the following 
will be present and non negligible due to the equivalence principle being already a good approximation for vector-like masses $M$ of 
the order of 600~GeV or more. However constraints  on different multiplet may be different and in some cases stronger (we indicated 
above flavor bounds, but also electroweak precision tests and low energy observables may give important bounds, 
see \cite{Cacciapaglia:2010vn,Cacciapaglia:2011fx,Okada:2012gy} for details).

\section{Production}
In this work we focus on the single production of $T$ in association with a light jet.
If only mixing to the third generation were allowed, a single $T$ would be produced uniquely in association with a top quark via an s-channel $Z$ boson, thus leading to typically small cross sections.
On the other hand, allowing a coupling to the up-quark, even if small, opens the possibility for single production in association with a light jet via s- and t-channel $Z$ exchanges, which offer much larger production rates~\cite{Buchkremer:2013bha}.

In order to study the production and hadronic decay of a singly produced heavy vector-like top, we have used standard Montecarlo 
simulation tools. The production of samples was done with MadGraph 5 \cite{Alwall:2011uj} version 1.5.11, for both signal and backgrounds. For the signal, the model was implemented using Feynrules \cite{Alloul:2013bka,Christensen:2008py}, using the model files in \cite{VLmodels}.
Finally, the hadronisation of the parton--level samples has been done with Pythia 6 \cite{Sjostrand:2006za}. These tools are well known to describe correctly high jet multiplicity final states, specifically an unstable particle (as the W, Z or t) with up to four additional jets. 
Cross sections and expected number of events for signal and each background involved in the analysis, at 8~TeV for 20~fb$^{-1}$, are listed in Table~\ref{tab:xsec}.

\begin{table}[tb]
\centering
\begin{tabular}{||l|c|r||}
  \hline\hline
  Process & $\sigma_{\rm 8 TeV}$ ($pb$) & Ex. Events \\ \hline
 Signal ($Tj$) & 0.2 & 700 \\
 \hline
  QCD (bbjjj) & 500 & 10,000,000 \\
  $W$+jets & 37,509 & 750,180,000 \\
  $Z$+jets & 3,503.71 & 70,074,200 \\ 
  $t\; \bar{t}$ & 234 & 4,680,000 \\
  single-$t$ & 114.85 & 2,297,000 \\
  Di-boson & 96.82 & 1,936,400 \\
  \hline\hline
\end{tabular}
\caption{Cross sections and number of events for signal and backgrounds (after generation cuts). Branching ratio to full hadronic final state was taken into account for the signal.} \label{tab:xsec}
\end{table} 

For the QCD sample, all jets were produced with a $p_{T}>30$~GeV and within a pseudo-rapidity of $|\eta|<5$. All the other background 
samples were produced with jets with $p_{T}>10$~GeV and no cut on the pseudo-rapidity. For all samples involving at least one Z, 
di--boson processes ZZ, WZ and Z+jets~\footnote{the mass of the di--lepton pair was required to satisfy $M_{ll}>50$~GeV, in order to avoid 
integration troubles}. In hadronisation step, the jets were built up with Anti-Kt algorithm with $R=0.5$ with the standard 
implementation provided by FastJet package \cite{Cacciari:2011ma}.

The signal sample was produced with $p_{T}>10$~GeV with the same packages. 
We chose to set the vector-like mass $M = 700$~GeV, and the mixing to their maximal allowed values when both mixing to third and first generation are allowed (and the coupling to the second generation is negligible) \cite{Buchkremer:2013bha}. With this choice, the physical mass of the $T$ is $M_{T}=734$~GeV, and around 700
events in full hadronic decay mode are expected at 8~TeV with 20~fb$^{-1}$. For this mass point the signal has a cross section around 
200~fb. The choice of the 
$T$ mass value is taken in the range expected to be accessible to 8~TeV LHC analyses in order to show the interest of performing a 
more detailed and dedicated analysis within the LHC collaborations. It is clear that analyses to be performed at a higher centre of mass 
energy in the future, for example at 13~TeV for the LHC and seeking particles in the TeV range, will require different dedicated analyses, 
for example using boosted objects~\cite{CMS:2013vca,TheATLAScollaboration:2013qia}. The preliminary study we perform will not cover this case as typical 
efficiencies are not known for a 13~TeV case and analysis techniques will be probably different.

\section{Analysis}
\label{sec:extract}
The final state we are interested in contains 5 jets (3 b-jets, two of which coming from a Higgs decay, and 2 jets from the $W$ decay) which come from the $T$, plus a sixth, more forward jet from the production.
The main issue with the analysis is to extract the signal over the large QCD background, as we aim to study a fully hadronic signal
for the benefit of having a larger signal cross-section times branching ratios and full mass reconstruction. All the different backgrounds 
with a similar signature were included in order to check their relevance in the analysis. Ordered by expected number of events, 
at $20$~fb$^{-1}$, and the difficulty to differentiate them from the signal the backgrounds involved were QCD, W $+$ jets, Z $+$ jets, 
$t\; \bar{t}$, single top, and di--boson ($WW$, $WZ$, $ZZ$). 

The fully hadronic channel allows the full mass reconstruction of the $T$. The analysis strategy will consist on reconstructing each decay products (top, Higgs) and then on looking for an accumulation of events around the $T$ mass while Standard Model background processes are expected to form a continuous distribution. Note also that this strategy is possible thanks to the relatively low mass range still allowed for the $T$, while for masses above the TeV (which will be accessible at the higher energy run of the LHC) the Higgs and W (top) are expected to be boosted and the jets from their decays merged.

First the jet association to reconstruct the decay products will be presented, then we will present the various criteria put in place in order to reduce background events.
The jets association was performed after basic kinematical selection but before any cut on the invariant masses.
The method used to identify the jets coming from the top partner is the following:
\begin{itemize}
\item Identify b--jets. (We only require at least two b's.)
\item For the Higgs decay, all pairs of b-jets passing the criteria $\Delta R_{jj} <2.5$ are retained. If more than one pair is  fulfilling the angular requirement, the pair with the closest mass to the Higgs mass (125~GeV) is chosen.
\item From the remaining jets, all possible pairs are formed and the one
with closest mass to W-mass (80~GeV) is identified as the W.
\item Finally, from the remaining jets, the previously identified W-jets are coupled with a third jet. The top b-jet is selected as the jet which, combined with W ones, gives the invariant mass closest to the top mass (172~GeV).
\end{itemize}

Characteristics of the signal have been exploited in order to differentiate it from the backgrounds. The main characteristic is the presence of a Higgs boson, as a handle to reduce the backgrounds. In Table \ref{tabxs} the cut flow to extract the signal is
shown together with the corresponding numbers of events expected for 20~fb$^{-1}$ in full hadronic decay mode and after the preliminary selection cuts. 
We do not quote numbers for the $Z$+jets background because our simulation contained a too low statistics to extract meaningful results (producing a $Z$ with a high jet multiplicity is computationally very expensive). Furthermore, the inclusive $Z$+jets cross section is one order of magnitude smaller than the $W$+jets one and, due to the similar branching ratios and kinematics, we expect similar efficiencies as on the $W$+jets background.
In the following we will therefore ignore this background, and assume that it can at most contribute an additional 10\% of the $W$+jets.
We also do not include the result for di--boson, as we checked that its contribution to the signal is negligible. 
One important remark is that the event selection was designed in order to not be altered, by much, when including the detector effects, this is, it was mostly based in angular distributions that are not greatly changed by detector simulation. In these terms, a big change on results is not expected when including the detector simulation. All cuts were applied one after the other in the order given in the following list:

\begin{itemize}

\item {\it Cut 0}: The first step consists in a preliminary event selection that mimics the case of a detector: an event is kept if it has at 
least 6 jets with $p_T > 30$~GeV, at least five jets within $|\eta|<2.5$ and at least one jet within $2<|\eta|<5$. The five jets coming 
from the $T$ are more central, while the sixth jet produced in association with the single $T$ tends to be forward. 
In Figure \ref{fig:Eta6thjet} we show the pseudo-rapidity of the sixth jet at parton level. 

\begin{figure}[h!]
\begin{center}
\includegraphics[scale=0.4, viewport=0 0 650 650, clip=true]{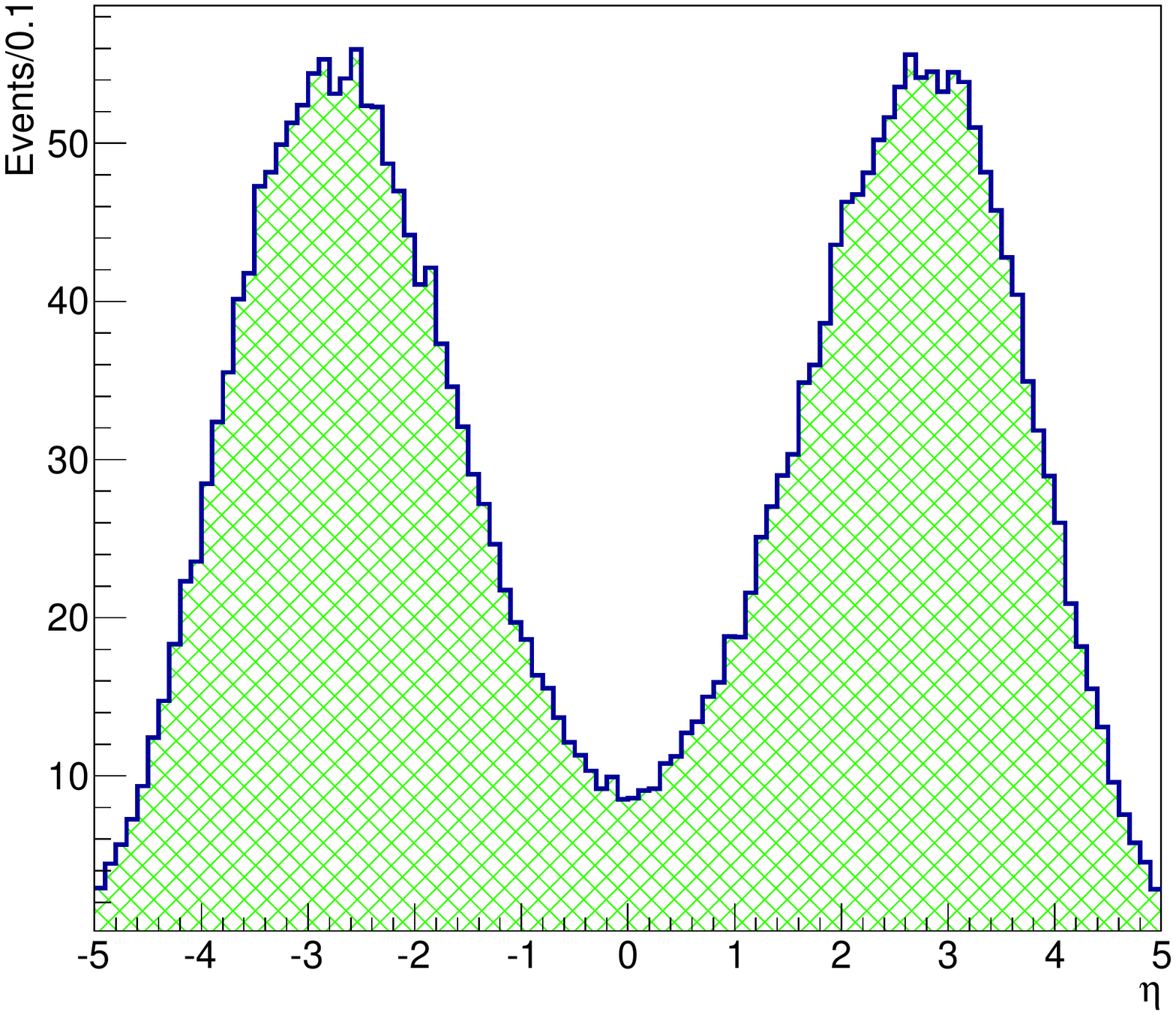}
\caption{\small \it  $\eta$ of the light jet that is produced in association with the top partner at parton level. 
\label{fig:Eta6thjet}}
\end{center}
\end{figure}

\begin{figure}[h!]
\begin{center}
\includegraphics[scale=0.77, viewport=0 0 700 750, clip=true]{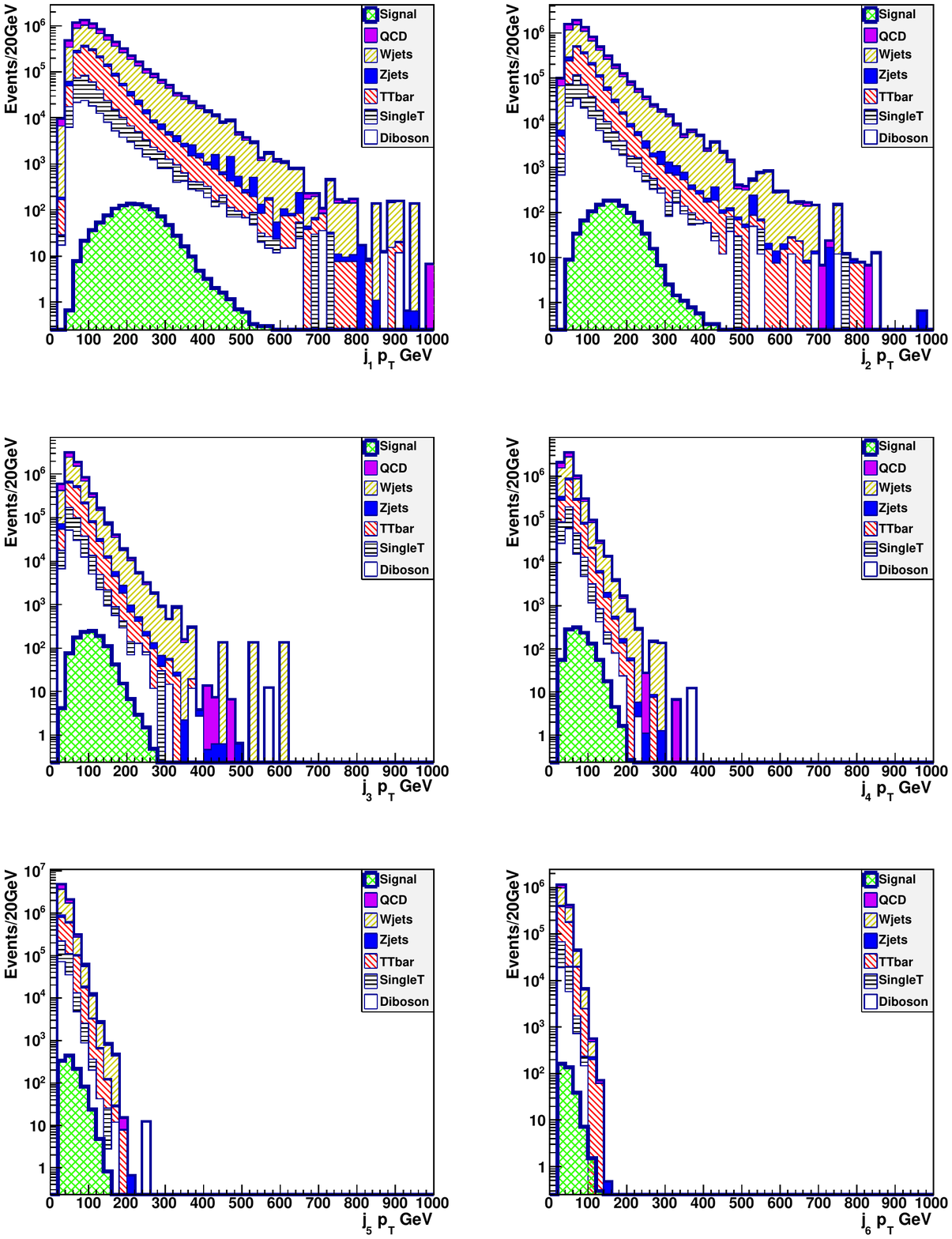}
\caption{\small \it  $p_{T}$  of the six leading jets for backgrounds (stacked) and signal (over--imposed). 
\label{fig:JetsPt}}
\end{center}
\end{figure}

\item {\it Cut 1}: First kinematical selection: $p_{T}>150$~GeV was required
for the leading jet in each event, $p_{T}>80$~GeV for the sub-leading one and $p_{T}>60$~GeV for the 3$^{rd}$ and 4$^{th}$ leading jets. Figure
\ref{fig:JetsPt} shows the distributions for the $p_{T}$ of the six leading jets for signal and backgrounds. 

\item {\it Cut 2}: The third cut involves the total hadronic energy ($H_{T}=\sum |p_{T}^{j}|$), which is plotted in Figure \ref{fig:THT} for backgrounds and signal. The massive $T$ is decaying into top and Higgs particles, which are the two heaviest particles of the Standard Model. This should lead to higher hadronic energy than for QCD, $W$ + jets
and $t$ $\bar{t}$ events. Events with $H_{T}>630$~GeV were selected.

\begin{figure}[h!]
\begin{center}
\includegraphics[scale=0.45, viewport=0 0 800 650, clip=true]{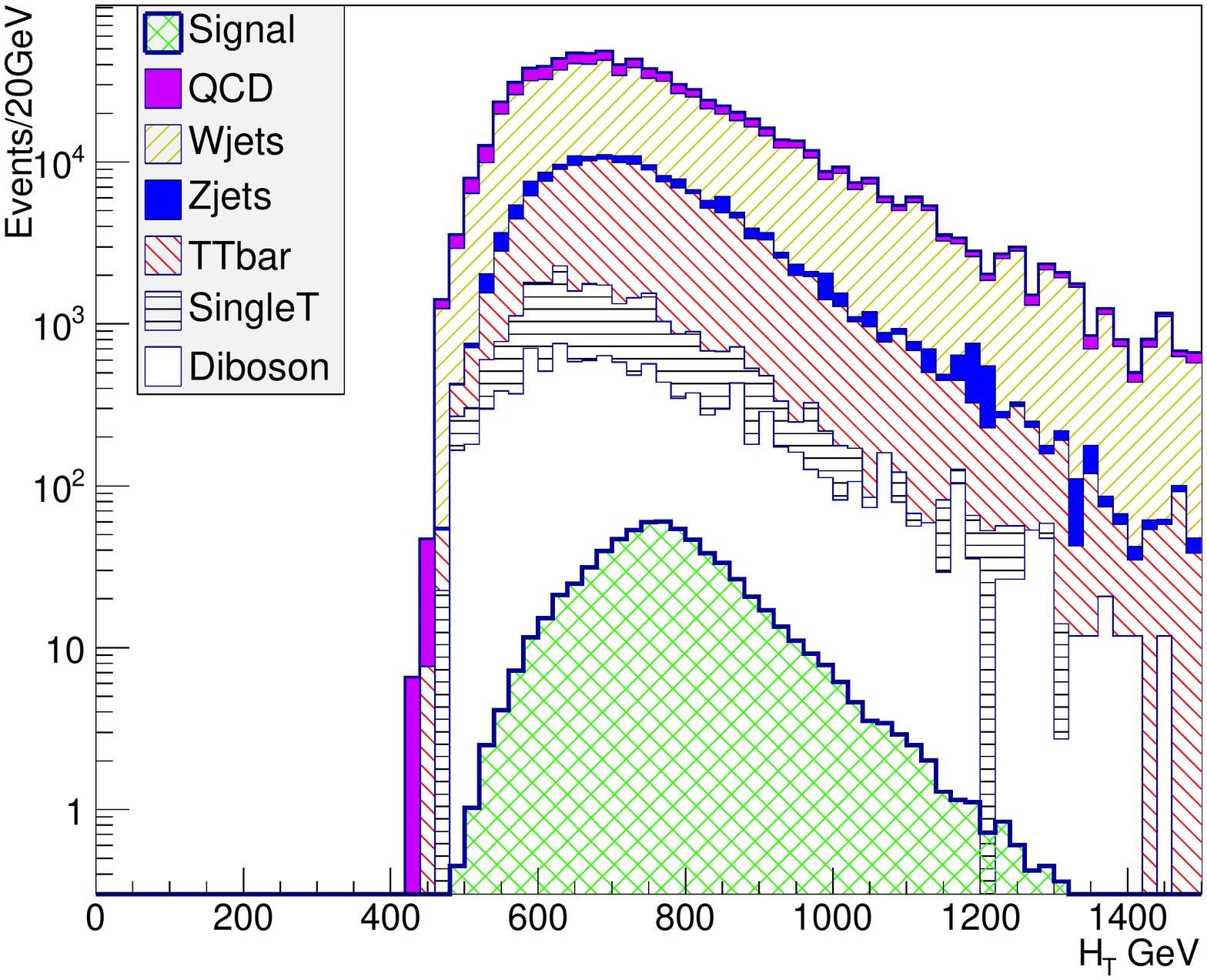}
\caption{\small \it  Total hadronic energy for backgrounds (stacked) and signal (over--imposed). 
\label{fig:THT}}
\end{center}
\end{figure}

\item {\it Cut 3}: We required each event to have at least two b jets. A very loose b tagging selection ($90\%$ efficient) was considered. A real performance study example of a b-tagging algorithm can be found in reference \cite{CMS:BTV}, where several working points are used for a displaced vertex algorithm in order to tag b-jets, for each working point the efficiency of the algorithm was determined in a jet by jet basis.

\item {\it Cut 4}: The jets coming from the Higgs boson have typically a $\Delta R_{jj}<1.8$. The Higgs boson is produced boosted within the decay of $T$. This non-zero momentum implies that the decay products of the Higgs boson tend to be close together.

\item {\it Cut 5}: The $p_{T}$ of the reconstructed Higgs and top quark for signal and backgrounds are shown in Figure \ref{fig:HptToppt},
in a 2D histogram. It is interesting to notice that in the case of backgrounds the reconstructed Higgs and top have a smaller $p_{T}$
than the signal. Only events which have a Higgs with $p_{T}>200$ and a top with $p_{T}>300$ were selected. 

\begin{figure}[h!]
\begin{center}
\includegraphics[scale=0.5, viewport=0 0 750 400, clip=true]{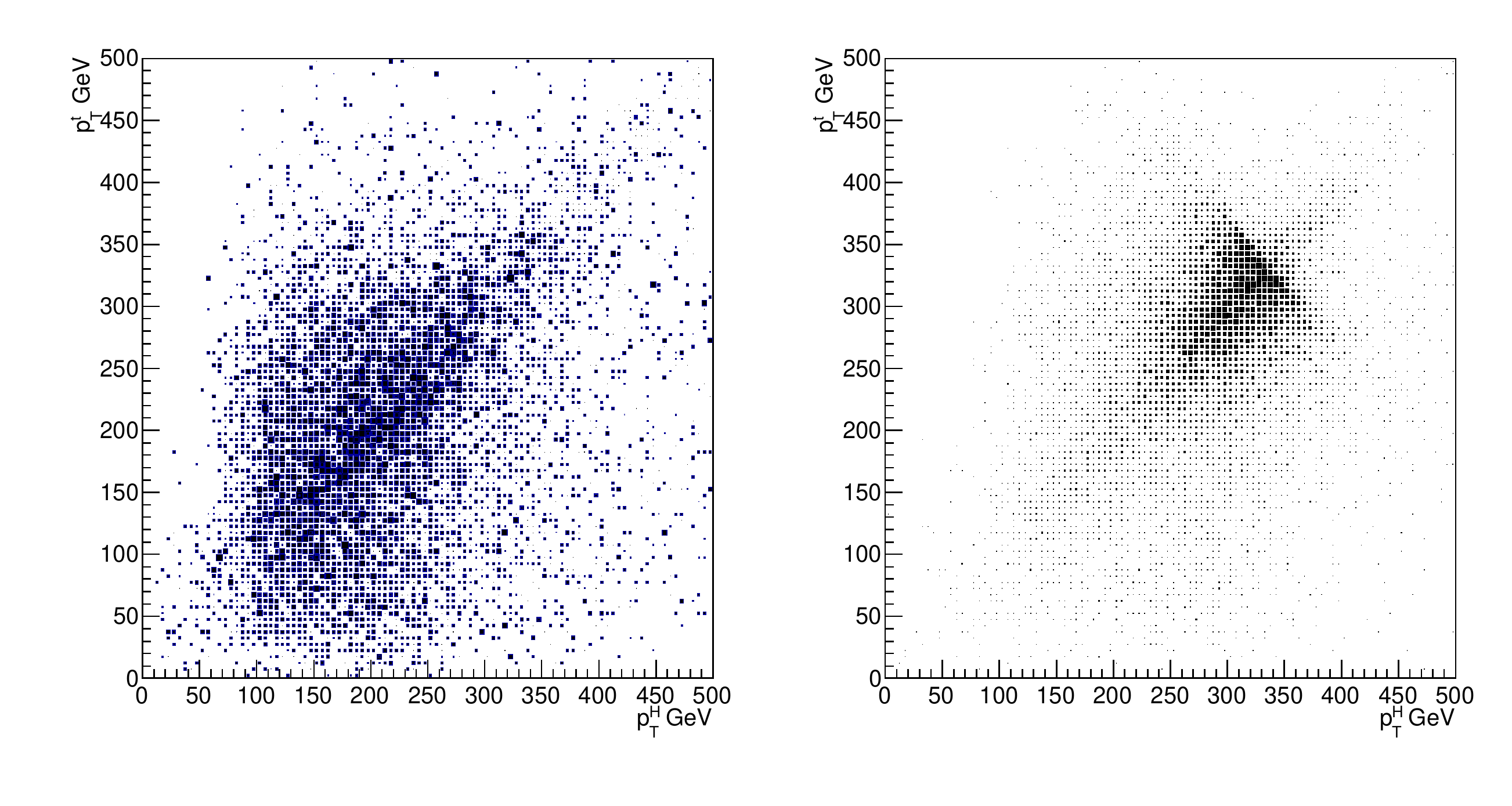}
\caption{\small \it Reconstructed Higgs $p_{T}$ in the x axis and reconstructed top $p_{T}$ in the y axis for backgrounds (left) and 
signal (right). 
\label{fig:HptToppt}}
\end{center}
\end{figure}

\item {\it Cut 6}: $\Delta R_{HW}$ is
the $\Delta R$ between the reconstructed Higgs and W, which is plotted in Figure \ref{fig:DRWH}. Selecting only the 
events within $2.2<\Delta R_{HW}<3.5$ helps to reduce QCD and W + jets background events.

\begin{figure}[h!]
\begin{center}
\includegraphics[scale=0.4, viewport=0 0 750 650, clip=true]{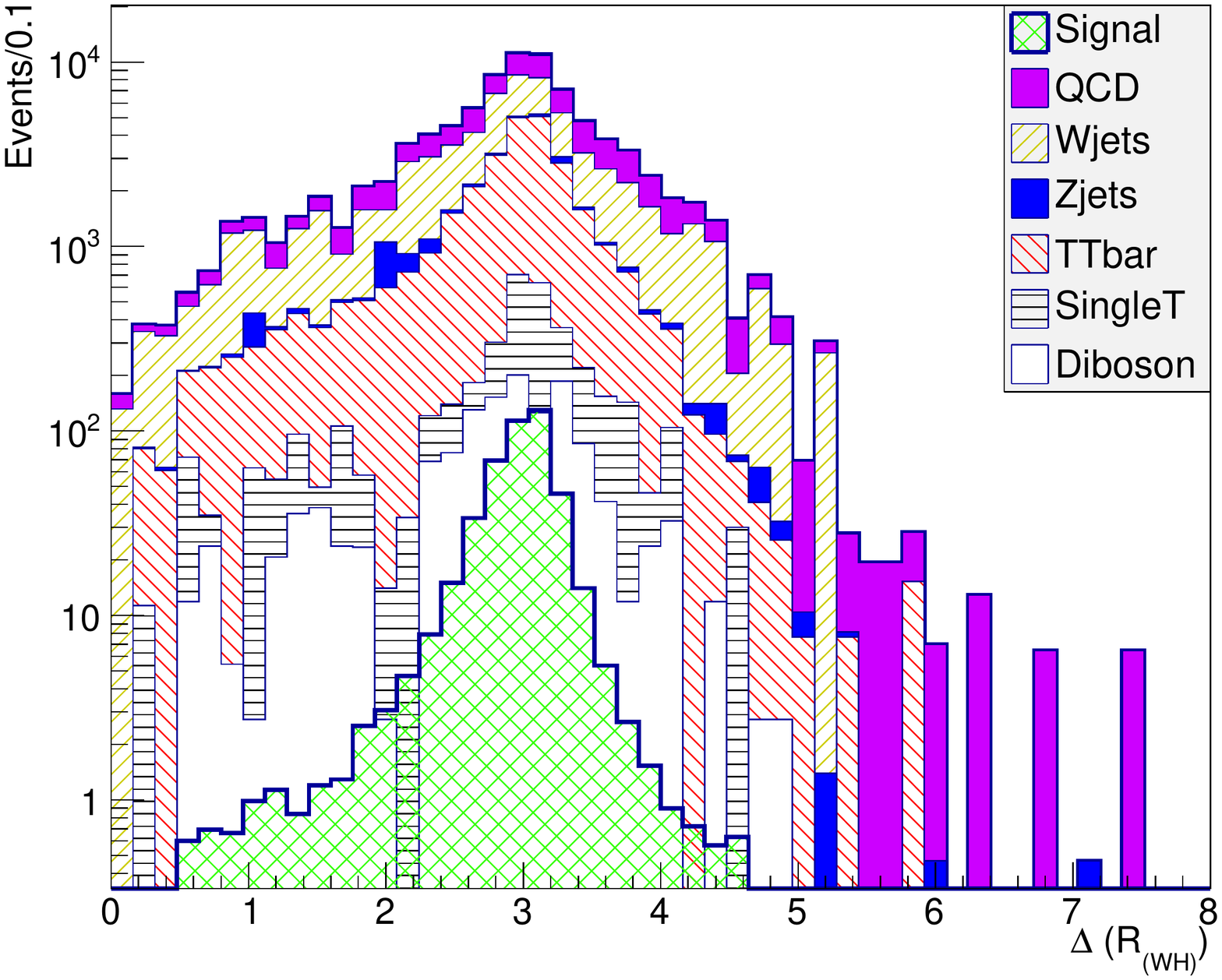}
\caption{\small \it $\Delta R$ between the reconstructed Higgs and W for backgrounds (stacked) and signal (over--imposed). 
\label{fig:DRWH}}
\end{center}
\end{figure}

\item {\it Cut 7}: The $\Delta \phi_{jj}$ of the b jets identified as coming from the Higgs boson and the $\Delta \phi_{jW}$ between the 
reconstructed W and the jet which formed the top are expected to be mainly central for the signal while more evenly distributed for 
backgrounds. The events 
that have $\Delta \phi_{jj}<2.0$ and $\Delta \phi_{jW}<3.3$ were selected. This cut is specially useful for reducing QCD and W+jets
background events.

\item {\it Cut 8}: The $\Delta \phi_{jj}$ between the jets of the W are also expected to be more 
centered around zero in the signal with respect to backgrounds. Only events with $\Delta \phi_{jj}<2.3$ were kept. This cut was required to reduce single--top background.

\item {\it Cut 9}: Finally, only events with a Higgs candidate with a mass between $100$~GeV and $135$~GeV 
were kept for the analysis.

\begin{figure}[h!]
\begin{center}
\includegraphics[scale=0.4, viewport=0 0 750 650, clip=true]{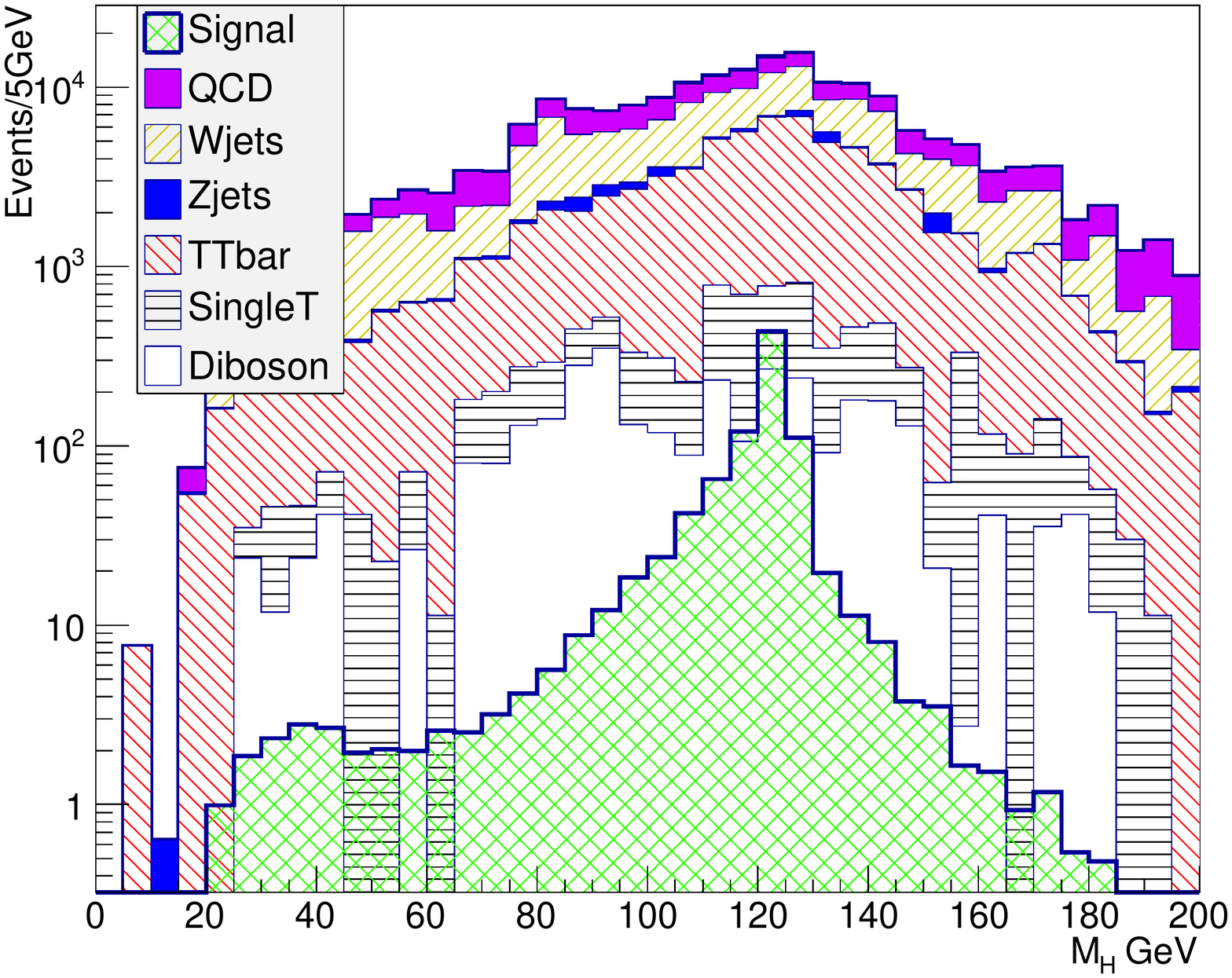}
\caption{\small \it Mass of the reconstructed Higgs for backgrounds (stacked) and signal (over--imposed). 
\label{fig:HiggsJetsMass}}
\end{center}
\end{figure}

\item {\it Cut 10}: An additional observable can be constructed in order to differentiate the signal from $t$ $\bar{t}$ background events.
It is the relative total hadronic energy, and it is defined as the ratio between the $p_{T}$ of the decay products identified as the Higgs
and top and the total hadronic energy of the event: $$\frac{p_{T}^{H}+p_{T}^{t}}{H_{T}}\,.$$ All events with a relative total
hadronic energy bigger than $0.65$ were kept. The corresponding plot is shown in
Figure \ref{fig:RelHT}.
\end{itemize}

\begin{table}[htb]
\centering
\begin{tabular}{l|c|c|c|c|c|c}
   & Signal & QCD & $W$+jets & $t \bar{t}$ & $t$+ jet & $tW$ \\ \hline
  Cut 0 & $909\pm 5$ & $203,930\pm 1,150$ & $1,015,294\pm 11,567$ &  $337,024\pm 1,608$ & $25,349\pm 300$ & $19,416\pm 469$ 
  \\ \hline
  Cut 1 & $0.91\pm 0.01$ & $0.571 \pm 0.007$ & $0.67\pm 0.02$ &  $0.439 \pm 0.005$ & $0.45 \pm 0.01$ & $0.42 \pm 0.03$ \\
  \hline
  Cut 2 & $0.92\pm 0.01$ &  $0.68\pm 0.01$ & $0.74 \pm 0.02$ &  $0.81 \pm 0.01$ & $0.61 \pm 0.02$ & $0.70 \pm 0.06$ \\
   \hline
  Cut 3 & $0.84\pm 0.01$ & $0.86 \pm 0.02$ & $0.22 \pm 0.01$ & $0.83 \pm 0.01$ & $0.82 \pm 0.04$ & $0.85 \pm 0.08$ \\
   \hline
  Cut 4 & $0.93\pm 0.01$ & $0.68 \pm 0.01$ & $0.74 \pm 0.06$ & $0.56 \pm 0.01$ & $0.49 \pm 0.03$ & $0.45 \pm 0.05$ \\
 \hline
  Cut 5 & $0.92\pm 0.01$ & $0.60 \pm 0.02$ & $0.56 \pm 0.05$ &  $0.53 \pm 0.01$ & $0.61 \pm 0.05$ & $0.56 \pm 0.09$ \\
  \hline
  Cut 6 & $0.92\pm 0.01$ & $0.61 \pm 0.02$ & $0.56 \pm 0.07$ &  $0.74 \pm 0.03$ & $0.66 \pm 0.07$ & $0.72 \pm 0.15$ \\
 \hline
  Cut 7 & $0.75\pm 0.01$ & $0.67 \pm 0.03$ & $0.67 \pm 0.11$ &  $0.71 \pm 0.03$ & $0.77 \pm 0.09$ & $0.75 \pm 0.18$ \\
  \hline
  Cut 8 & $0.87\pm 0.02$ & $0.76 \pm 0.04$ & $0.82 \pm 0.15$ &  $0.84 \pm 0.04$ & $0.77 \pm 0.11$ & $0.90 \pm 0.24$ \\
\hline
  Cut 9 & $0.91\pm 0.02$ & $0.33 \pm 0.02$ & $0.41 \pm 0.10$ &  $0.51 \pm 0.03$ & $0.52 \pm 0.09$ & $0.48 \pm 0.16$ \\
  \hline
  Cut 10 & $0.87\pm 0.02$ & $0.54 \pm 0.06$ & $0.55 \pm 0.19$ &  $0.49 \pm 0.04$ & $0.79 \pm 0.17$ & $0.72 \pm 0.31$ \\
\hline
  combined & $0.284\pm 0.005$ & $(7.5 \pm 0.5) \times 10^{-3}$ & $(3.1 \pm 0.7) \times 10^{-3}$ &  $(9.3 \pm 0.5) \times 10^{-3}$ & $(11 \pm 1) \times 10^{-3}$ & $(10.5 \pm 3) \times 10^{-3}$ \\
 \end{tabular}
\caption{Number of events for signal and backgrounds after the first selection cut (Cut 0), and efficiencies of each stage of the cutting procedure. The errors indicated are statistical only, based on the number of events.} \label{tab:nev}
\label{tabxs}
\end{table}

\begin{figure}[h!]
\begin{center}
\includegraphics[scale=0.4, viewport=0 0 700 650, clip=true]{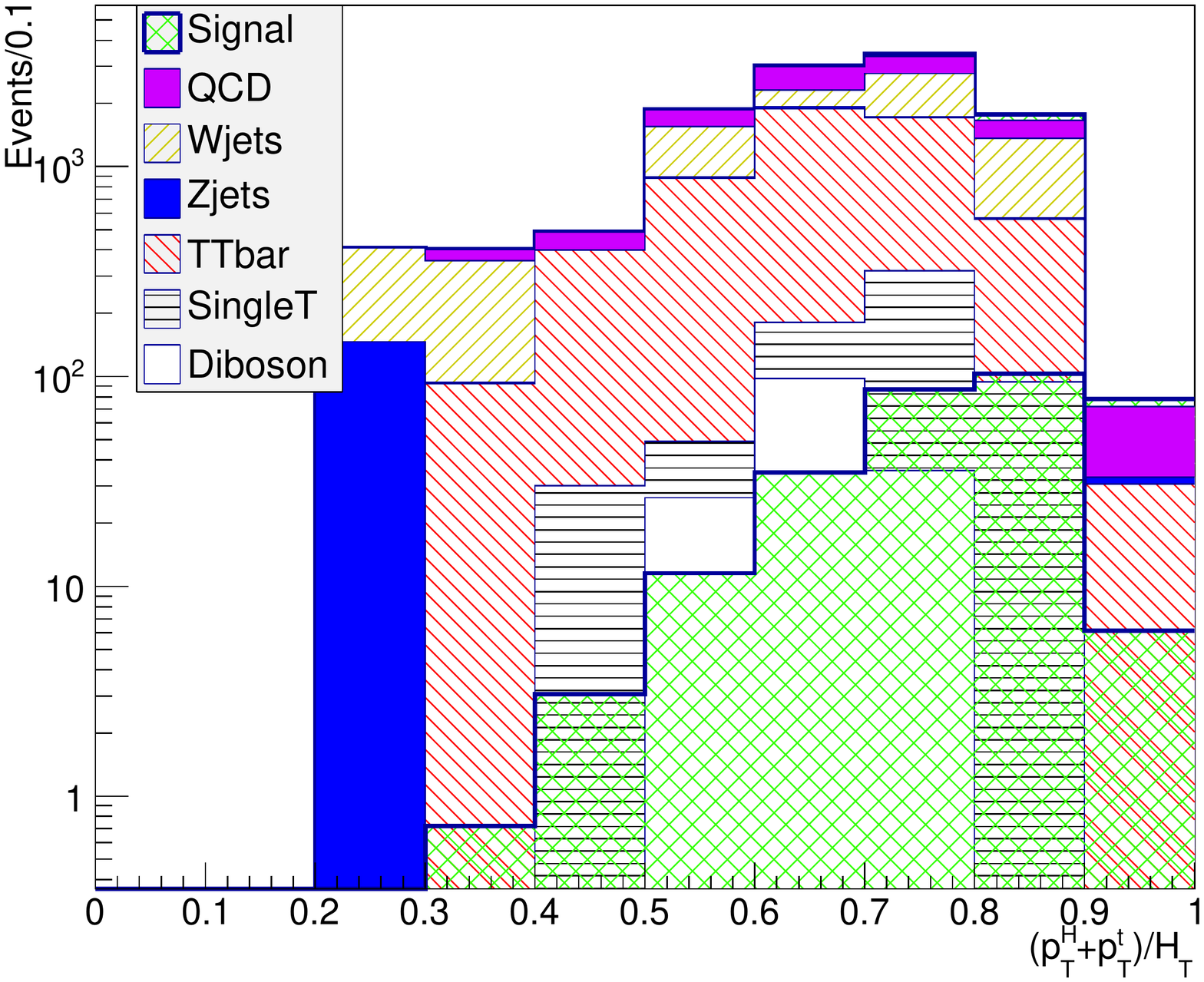}
\caption{\small \it Relative total hadronic energy for backgrounds (stacked) and signal (over--imposed). 
\label{fig:RelHT}}
\end{center}
\end{figure}

In Table~\ref{tab:nev}, we show the efficiency of each cut in the procedure described above: the first line contains the number of events 
after the initial Cut 0 normalised to an integrated luminosity of 20~fb$^{-1}$, while on the following lines the efficiencies of the cuts for signal and background samples are shown after each of 
the 10 cuts. The bottom line of the table contains combined efficiency of all the cuts on each sample of events. 
As this analysis was done at the hadronic level (after showering), more jets were expected in comparison to parton level due to 
radiation, added by the showering, that may be accidentally reconstructed as an additional jet by the jet reconstruction algorithm and is not associated with a parton.
 
\begin{figure}[h!]
\begin{center}
\includegraphics[scale=0.4, viewport=0 0 700 650, clip=true]{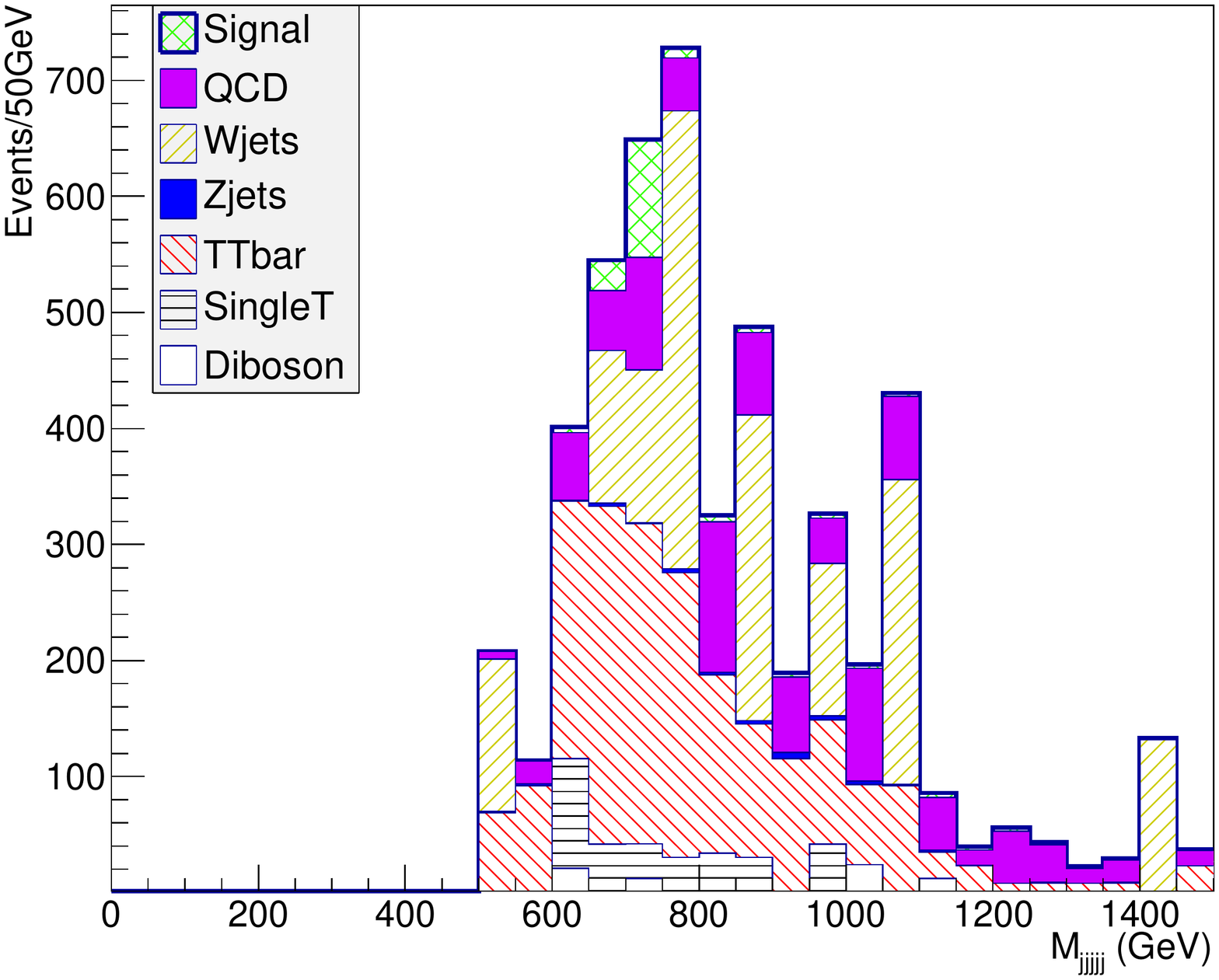}
\caption{\small \it  Reconstructed T mass after all cuts for backgrounds and signal (stacked). 
\label{fig:AfterCuts}}
\end{center}
\end{figure}

In Figure \ref{fig:AfterCuts} the peak reconstruction is shown, with the aforementioned identification procedure and after all the cuts. 
In this figure a clear peak around 730~GeV is observed with the signal clearly visible over the ensemble of backgrounds. 
The lack of smoothness of the distribution is due to the lack of statistics in the MonteCarlo samples for the 
backgrounds, especially for W+jets. These fluctuations due to statistics can change the final estimate of the number of background 
events entering the peak of the signal, and the error is partially accounted for in the statistical error. After all cuts, we selected the number of events falling into a window of $20$~GeV around 
the $T$ mass, i.e. within $710 < M_{jjjjj} < 750$~GeV. The number of events we obtain are listed in Table~\ref{tab:events}. We therefore obtain an enhanced signal over background ratio, with:
\begin{equation}
\frac{S}{\sqrt{S+B}}=3.2\pm 0.6\,, \qquad \mbox{and} \quad \frac{S}{B}=0.10\pm 0.03\,. 
\end{equation}
To the quoted uncertainties, one should add uncertainties in the cross section calculation for the signal, on PDF's and possible 
loop contributions. From similar studies and in analysis done by CMS and ATLAS collaborations the uncertainties linked to these sources are not bigger than 10\% to 15\% (see for instance \cite{Aad:2011yn}), therefore their inclusion should not change significantly the conclusions of this study.

\begin{table}[tb]
\begin{center}
\begin{tabular}{l|c|c|c|c}
 & \multicolumn{2}{c|}{unweighted events}  & \multirow{2}{*}{weight} & weighted  \\
 & after Cut 10 & in mass window & & events \\
 \hline
 Signal & $8601$ & $3780$ & $0.03$ &$113 \pm 2$ \\
 \hline
   $t \bar{t}$ & $409$ & $57$ & $7.7$ & $437 \pm 58$ \\
 $W$+jets & $24$ & $3$ & $132$ & $395 \pm 228$ \\
 $QCD$ & $235$ & $34$ & $6.48$ & $220 \pm 38$ \\
 $tW$ & $18$ & $3$ & $11.3$ & $34 \pm 20$ \\
 $t$+ jet & $75$ & $7$ & $3.55$ & $25 \pm 9$ \\
  \hline
  total background & & & & $1112 \pm 352$ \\
\end{tabular}
\caption{Number of signal and background events from our simulation: in the first column the simulated events that pass all kinematic cuts, in the second column the events that fall in the mass window $710 < M_{jjjjj} < 750$~GeV, finally in the fourth column the number of weighted events in the mass window normalized to the physical cross section (the applied weight is listed in the third column). All the errors are statistical only. For the total background, we conservatively consider linear sum of errors.} \label{tab:events} \end{center}
\end{table}

The analysis strategy based simulation was performed with a physical width of the $T$ particle preset to $1$~GeV in the FeynRules implementation. The couplings we used, however, give a physical width of $11$~GeV. In general, it is interesting to test how the result we quote may be affected by the physical width of the $T$, which is directly sensitive to the size of the couplings. Changing the physical width does not affect any of the selection criteria, but simply affects the final selection based on the invariant mass of the 5 jets. We checked that, with a width of $11$~GeV, an opening up to $30$~GeV as mass window is necessary to include around 2 sigma of the signal. The $S/\sqrt{S+B}$ does not change significantly within the statistical errors in our analysis. We can therefore conclude that the physical width of the $T$ does not affect significantly this search. 
This study was done with the help of MadAnalysis5 package \cite{Conte:2012fm}.

\section{Conclusions}
\label{sec:con}

We have performed a simplified analysis for single production of a vector-like top partner $T$ in association with a light jet, with the $T$ decaying into a top quark and a Higgs boson 
in the fully hadronic channel.
We thus demonstrated a 
strategy to extract the signal over the background, considering as a benchmark the 20~fb$^{-1}$ of data at 8~TeV collected within the 
latest run of the LHC and a benchmark mass around $730$~GeV, which is within the range of reach for these analyses. This study suggests the interest in 
performing a dedicated analysis by the LHC experimental collaborations. 
Indeed the larger branchings in the hadronic decays give a higher number of signal events and full mass reconstruction is 
achievable, something which is not possible in the corresponding leptonic signatures studied at present.

This study also shows the importance of even small mixings of the $T$ to the first generation as production is driven by these 
first generation couplings due to the large light partonic content in the colliding protons. The couplings to the third 
generation and in particular to the top quark, allow decay modes which give clear signatures with respect to background 
processes from the standard model.

\section*{Acknowledgments}
We would like to thank the MadAnalysis team for their technical support, in particular E.~Conte and B.~Fuks.
AD is partially supported by Institut Universitaire de France. We also acknowledge partial support from 
the Labex-LIO (Lyon Institute of Origins) under grant ANR-10-LABX-66 and FRAMA (FR3127, F\'ed\'eration de Recherche ``Andr\'e 
Marie Amp\`ere").


\begin{thebibliography}{99}

\bibitem{delAguila:1982fs}
  F.~del Aguila and M.~J.~Bowick,
  Nucl.\ Phys.\ B {\bf 224} (1983) 107.
  
\bibitem{Aad:2012uu}
  G.~Aad {\it et al.}  [ATLAS Collaboration],
  JHEP {\bf 1211} (2012) 094
  [\href{http://arxiv.org/abs/1209.4186}{arXiv:1209.4186} [hep-ex]].

\bibitem{ATLAS:2013ima}
  [ATLAS Collaboration],
  \href{http://cds.cern.ch/record/1525525?ln=en}{ATLAS-CONF-2013-018}
  
\bibitem{Aad:2011yn}
  G.~Aad {\it et al.}  [ATLAS Collaboration],
  CERN-PH-EP-2011-193
  Phys.\ Lett.\ B {\bf 712} (2012) 22
  [\href{http://arxiv.org/abs/1112.5755}{arXiv:1112.5755} [hep-ex]].
  
\bibitem{Aad:2012bt}
  G.~Aad {\it et al.}  [ATLAS Collaboration],
  CERN-PH-EP-2012-008
  Phys.\ Rev.\ D {\bf 86} (2012) 012007
  \href{http://arxiv.org/abs/1202.3389}{arXiv:1202.3389} [hep-ex]].
  
\bibitem{CMSincl}
  S.~Chatrchyan {\it et al.}  [CMS Collaboration], 
 \href{http://cds.cern.ch/record/1557571?ln=en}{CMS-PAS-B2G-12-015}

\bibitem{Chatrchyan:2012af}
  S.~Chatrchyan {\it et al.}  [CMS Collaboration],
  JHEP {\bf 1301} (2013) 154
    [\href{http://arxiv.org/abs/1210.7471}{arXiv:1210.7471} [hep-ex]].

\bibitem{Chatrchyan:2012fp}
  S.~Chatrchyan {\it et al.}  [CMS Collaboration],
  Phys.\ Rev.\ D {\bf 86} (2012) 112003
      [\href{http://arxiv.org/abs/1209.1062}{arXiv:1209.1062} [hep-ex]].
  
\bibitem{Chatrchyan:2012vu}
  S.~Chatrchyan {\it et al.}  [CMS Collaboration],
  Phys.\ Lett.\ B {\bf 718} (2012) 307
        [\href{http://arxiv.org/abs/1209.0471}{arXiv:1209.0471} [hep-ex]].
  
\bibitem{CMS:2012ab}
  S.~Chatrchyan {\it et al.}  [CMS Collaboration],
  Phys.\ Lett.\ B {\bf 716} (2012) 103
          [\href{http://arxiv.org/abs/1203.5410}{arXiv:1203.5410} [hep-ex]].
  
\bibitem{Chatrchyan:2011ay}
  S.~Chatrchyan {\it et al.}  [CMS Collaboration],
  Phys.\ Rev.\ Lett.\  {\bf 107} (2011) 271802
            [\href{http://arxiv.org/abs/1109.4985}{arXiv:1109.4985} [hep-ex]].
  
\bibitem{Buchkremer:2013bha}
  M.~Buchkremer, G.~Cacciapaglia, A.~Deandrea and L.~Panizzi,
  Nucl.\ Phys.\ B {\bf 876} (2013) 376
       [\href{http://arxiv.org/abs/1305.4172}{arXiv:1305.4172} [hep-ph]].

\bibitem{Aguilar-Saavedra:2013qpa}
  J.~A.~Aguilar-Saavedra, R.~Benbrik, S.~Heinemeyer and M.~Perez-Victoria,
  Phys.\ Rev.\ D {\bf 88} (2013) 094010
    [\href{http://arxiv.org/abs/1306.0572}{arXiv:1306.0572} [hep-ph]].

\bibitem{delAguila:2000rc}
  F.~del Aguila, M.~Perez-Victoria and J.~Santiago,
  JHEP {\bf 0009} (2000) 011
 \href{http://arxiv.org/abs/hep-ph/0007316}{[hep-ph/0007316]}.
  
\bibitem{ATLAS:2012apa}
  [ATLAS Collaboration],
    \href{http://cds.cern.ch/record/1480628}{ATLAS-CONF-2012-137}.
   
\bibitem{Garberson:2013jz}
  F.~Garberson and T.~Golling,
  Phys.\ Rev.\ D {\bf 87} (2013) 7,  072007
    [\href{http://arxiv.org/abs/1301.4454}{arXiv:1301.4454} [hep-ex]].
  
\bibitem{Li:2013xba}
  J.~Li, D.~Liu and J.~Shu,
  JHEP {\bf 1311} (2013) 047
  [arXiv:1306.5841 [hep-ph]].
  
\bibitem{Cacciapaglia:2010vn}
  G.~Cacciapaglia, A.~Deandrea, D.~Harada and Y.~Okada,
  JHEP {\bf 1011} (2010) 159
  [\href{http://arxiv.org/abs/1007.2933}{arXiv:1007.2933} [hep-ph]].
  
\bibitem{Cacciapaglia:2011fx}
  G.~Cacciapaglia, A.~Deandrea, L.~Panizzi, N.~Gaur, D.~Harada and Y.~Okada,
  JHEP {\bf 1203} (2012) 070
   [\href{http://arxiv.org/abs/1108.6329}{arXiv:1108.6329} [hep-ph]].
  
\bibitem{Okada:2012gy}
  Y.~Okada and L.~Panizzi,
  Adv.\ High Energy Phys.\  {\bf 2013} (2013) 364936
     [\href{http://arxiv.org/abs/1207.5607}{arXiv:1207.5607} [hep-ph]].

\bibitem{Cacciapaglia:2012dd}
  G.~Cacciapaglia, A.~Deandrea, L.~Panizzi, S.~Perries and V.~Sordini,
  JHEP {\bf 1303} (2013) 004
  [arXiv:1211.4034 [hep-ph]].

\bibitem{Azatov:2013hya}
  A.~Azatov, M.~Salvarezza, M.~Son and M.~Spannowsky,
  Phys.\ Rev.\ D {\bf 89} (2014) 075001
  [arXiv:1308.6601 [hep-ph]].

\bibitem{Contino:2008hi}
  R.~Contino and G.~Servant,
  JHEP {\bf 0806} (2008) 026
  [arXiv:0801.1679 [hep-ph]].

\bibitem{Mrazek:2009yu}
  J.~Mrazek and A.~Wulzer,
  Phys.\ Rev.\ D {\bf 81} (2010) 075006
  [arXiv:0909.3977 [hep-ph]].
  
\bibitem{Dissertori:2010ug}
  G.~Dissertori, E.~Furlan, F.~Moortgat and P.~Nef,
  JHEP {\bf 1009} (2010) 019
  [arXiv:1005.4414 [hep-ph]].

  
\bibitem{DeSimone:2012fs}
  A.~De Simone, O.~Matsedonskyi, R.~Rattazzi and A.~Wulzer,
  JHEP {\bf 1304} (2013) 004
  [arXiv:1211.5663 [hep-ph]].


\bibitem{Gopalakrishna:2013hua}
  S.~Gopalakrishna, T.~Mandal, S.~Mitra and G.~Moreau,
  arXiv:1306.2656 [hep-ph].
  

  
\bibitem{ATLAS:2012qe}
  G.~Aad {\it et al.}  [ATLAS Collaboration],
  Phys.\ Lett.\ B {\bf 718} (2013) 1284
       [\href{http://arxiv.org/abs/1210.5468}{arXiv:1210.5468} [hep-ex]].
  

 \bibitem{ATLAS:2013-051}
  [ATLAS Collaboration],
  \href{http://cds.cern.ch/record/1547567?ln=en}{ATLAS-CONF-2013-051}
    
\bibitem{Alwall:2011uj}
  J.~Alwall, M.~Herquet, F.~Maltoni, O.~Mattelaer and T.~Stelzer,
  JHEP {\bf 1106} (2011) 128
       [\href{http://arxiv.org/abs/1106.0522}{arXiv:1106.0522} [hep-ph]].
  
\bibitem{Alloul:2013bka}
  A.~Alloul, N.~D.~Christensen, C.~Degrande, C.~Duhr and B.~Fuks,
       [\href{http://arxiv.org/abs/1310.1921}{arXiv:1310.1921} [hep-ph]].

\bibitem{Christensen:2008py}
  N.~D.~Christensen and C.~Duhr,
  [\href{http://arxiv.org/abs/0806.4194}{arXiv:0806.4194} [hep-ph]]. 
  
\bibitem{VLmodels}
 \href{http://feynrules.irmp.ucl.ac.be/wiki/VLQ}{http://feynrules.irmp.ucl.ac.be/wiki/VLQ}

\bibitem{Sjostrand:2006za}
  T.~Sjostrand, S.~Mrenna and P.~Z.~Skands,
  JHEP {\bf 0605} (2006) 026
   \href{http://arxiv.org/abs/hep-ph/0603175}{[hep-ph/0603175]}.


\bibitem{Cacciari:2011ma}
  M.~Cacciari, G.~P.~Salam and G.~Soyez,
  Eur.\ Phys.\ J.\ C {\bf 72} (2012) 1896
  [arXiv:1111.6097 [hep-ph]].
      



\bibitem{CMS:2013vca}
  CMS Collaboration [CMS Collaboration],
  CMS-PAS-B2G-12-005.

      
\bibitem{TheATLAScollaboration:2013qia}
  The ATLAS collaboration,
  ATLAS-CONF-2013-084.

\bibitem{CMS:BTV}
  CMS Collaboration [CMS Collaboration],
  CMS-PAS-BTV-13-001.

\bibitem{Conte:2012fm}
  E.~Conte, B.~Fuks and G.~Serret,
  Comput.\ Phys.\ Commun.\  {\bf 184} (2013) 222
  [arXiv:1206.1599 [hep-ph]].

\end{thebibliography}
\end{document}